\documentclass{webofc}
\usepackage[varg]{txfonts}
\usepackage{slashed}
\usepackage{epsfig,latexsym,cancel,amssymb,amsmath,verbatim,mathrsfs}
\usepackage{color}
\usepackage{graphicx}
\def\ra{\rightarrow}
\def\L{\left(}
\def\R{\right)}
\def\wt{\widetilde}
\def\ld{\lambda}
\def\f{\frac}
\newcommand{\GeV}{\text{GeV}}
\newcommand{\eff}{\text{eff}}

\begin{document}

\fancyhead[C]{\it {
The proceedings of Joint Conference of ICGAC-XIII and IK15, Korea, 3--7 July 2017
}}
\fancyfoot[C]{\thepage}

\title{
Gravitational waves from the first order electroweak phase transition in the $Z_3$ symmetric singlet scalar model~\footnote{This proceeding paper is based on Ref.~\cite{Kang:2017mkl} in collaborated with Zhaofeng Kang and Pyungwon Ko.}
}
\author{\firstname{Toshinori} \lastname{Matsui}\inst{1}\fnsep\thanks{\email{matsui@kias.re.kr}}}
\institute{School of Physics, KIAS, Seoul 02455, Korea}
\abstract{
 Among various scenarios of baryon asymmetry of the Universe, electroweak baryogenesis is directly connected with physics of the Higgs sector.
 We discuss spectra of gravitational waves which are originated by the strongly first order phase transition at the electroweak symmetry breaking, which is required for a successful scenario of electroweak baryogenesis.
 In the $Z_3$ symmetric singlet scalar model, the significant gravitational waves are caused by the multi-step phase transition. 
 We show that the model can be tested by measuring the characteristic spectra of the gravitational waves at future interferometers such as LISA and DECIGO. 
}
\maketitle

%%%%%%%%%%%%%%%%%%%%%%%%%%%%
%%%%%%%%%%%%%%%%%%%%%%%%%%%%
%%%%%%%%%  sec: introduction   %%%%%%%%%
%%%%%%%%%%%%%%%%%%%%%%%%%%%%
%%%%%%%%%%%%%%%%%%%%%%%%%%%%
\section{Introduction}\label{sec:intro}

 In the scenario of electroweak baryogenesis (EWBG)~\cite{EWBG,Morrissey:2012db}, the strongly first order electroweak phase transition (SFOEWPT) is required to satisfy the condition of the departure from thermal equilibrium
%%----- eqnarray >>>>>
\begin{eqnarray}\label{eq:SFOEWPT}
\langle h\rangle_*/T_*\gtrsim 1,
\end{eqnarray}
%%<<<<< eqnarray -----
with $T_*$ being the temperature of EWPT and $\langle h\rangle_*$ the vacuum expected value (VEV) of the SM Higgs field $h$ at $T_*$.
 In order to satisfy this condition, the extended Higgs sector from standard model (SM) is required. 
 These extensions could help to build a barrier between the EW vacuum and a metastable vacuum at tree or loop level~\cite{Morrissey:2012db,Chung:2012vg}.
 The mechanism to generate a thermal cubic term for $h$ by a tree level barrier is most easily implemented in the extended Higgs sectors by a singlet $S$, containing effective tree-level cubic terms $\sim S^3+S|H|^2$ with $H$ the SM Higgs doublet~\cite{Profumo:2007wc,Ashoorioon:2009nf,2step1,Fuyuto:2014yia,Profumo:2014opa,Huang:2016cjm,Hashino:2016xoj}. 

 If the extended Higgs sector respects some symmetry such as $Z_2$, under which $S\ra -S$ and $H\ra H$, an alternative way to the desired tree level barrier is available in the symmetric limit where $S$ does not acquire VEV at the present universe~\cite{2step1,2step2,Curtin:2014jma,Vaskonen:2016yiu,Beniwal:2017eik,Kurup:2017dzf}.
 Such a scenario is associated with multi-step PT's.
 The universe may have been once in the intermediate phase $\Omega_{\rm meta}$ and then tunneled through a tree level barrier to the phase $\Omega_{\rm EW}$, recovering the $Z_2$ symmetry.

 We expect that gravitational wave (GW) is available to explore the nightmare scenario which is a case that the model cannot be tested at colliders.
 In principle, EWPT of $T_*\simeq 100~\GeV$ can be detectable at the GW observation experiments~\cite{Caprini:2015zlo}.
 The space-based interferometers: LISA~\cite{Seoane:2013qna}, DECIGO~\cite{Kawamura:2011zz} and BBO~\cite{Corbin:2005ny}, designed to be sensitive to GW density $\Omega_{\rm GW} h^2\gtrsim 10^{-16}-10^{-10}$ (depending on frequency $\simeq 10^{-3}-10^{-1}$ Hz), will be launched in the near future~\cite{Caprini:2015zlo}.

%%%%%%%%%%%%%%%%%%%%%%%%%%%%
%%%%%%%%%%%%%%%%%%%%%%%%%%%%
%%%%%%%%%  sec:model    %%%%%%%%%
%%%%%%%%%%%%%%%%%%%%%%%%%%%%
%%%%%%%%%%%%%%%%%%%%%%%%%%%%
\section{$Z_3$ symmetric singlet scalar model}\label{sec:model}

 We introduce an isospin complex singlet scalar $S$ transforming as $S\ra e^{i2w} S$ with $w=\pi/3$ under $Z_3$, while the SM fields including the SM Higgs doublet $H$ are neutral under $Z_3$.
 The most general renormalizable and $Z_3$-symmetric scalar potential $V(H,S)$ is given by
%%----- eqnarray >>>>>
\begin{eqnarray} %\label{}
V_0(H,S)=-\mu_h^2|H|^2-\mu_s^2|S|^2+{\ld_h}|H|^4+ \ld_s|S|^4 +\ld_{sh}|H|^2|S|^2 +\sqrt{2}\L \f{A_s}{3} S^3+h.c.\R. 
\end{eqnarray}
%%<<<<< eqnarray -----
 Compared to the $Z_2$-symmetric model, there is just one more parameter describing the cubic term $A_sS^3$.~\footnote{
 In this paper we do not consider the possibility that $S$ makes the dark matter (DM) candidate~\cite{Belanger:2012zr}, because we failed in finding viable parameter space with $\ld_{sh}\sim {\cal O}(0.01)$ that is necessary to accommodate correct DM phenomenology.
}
 After EWSB, two scalar fields are parametrized as $H=(G^+, (v+h^0+i G^0)/\sqrt{2})$ and $S=(s^0+ia_s^0)/\sqrt{2}$.
 There appear two physical degrees of freedom $h$ and $s$ in addition to Nambu-Goldstone (NG) modes $G^\pm$ and $G^0$ that are absorbed by the W- and Z-bosons.
 The vacuum stability condition reads as $\ld_s>0$, $\ld_h>0$ and $4\ld_s\ld_h>\ld_{sh}^2$. At zero temperature $T=0$, the model parameters are fixed to be $\ld_h= m_h^2/(2v^2)$, $\mu_h^2 = m_h^2/2$ and $\mu_s^2 = \ld_{sh} v^2/2-m_s^2$ up to radiative corrections with $v$ which is the VEV of $h$. Here, $m_h$ and $m_s$ are the physical masses of $h$ and $s$.
 We use $v=246~\GeV, m_h=125~\GeV, m_s, \ld_s, \ld_{sh}$ and $A_s$ as the input parameters. 

 Expanding the scalar fields around their classical backgrounds, $\left\langle H \right\rangle=(0, {\varphi_h}/\sqrt{2})$ and $\left\langle S \right\rangle={\varphi_s}/\sqrt{2}$, the one-loop effective potential at finite temperature is given by
%%----- eqnarray >>>>>
\begin{eqnarray} \label{eq:effpot}
\scalebox{0.82}{$\displaystyle
  V_{\eff}^{}(\varphi_h^{},\varphi_s^{}, T)
  =
V_0(\left\langle H \right\rangle,\left\langle S \right\rangle)
  +\sum_i n_i^{} \ 
  \f{M^4_i(\varphi_h^{},\varphi_s^{}, T)}{64\pi^2}
  \L\ln\f{M^2_i(\varphi_h^{},\varphi_s^{}, T)}{Q^2} -c_i \R + \sum_i n_i \ \f{T^4}{2\pi^2}I_{B,F} 
  \L \f{ M^2_i(\varphi_h^{},\varphi_s^{}, T)}{T^2} \R, 
  $}
\end{eqnarray}
%%<<<<< eqnarray -----
where $Q$ is the renormalization scale, which is set at $v^{}$ in our analysis.
 Here, $n_i$ and $M_i(\varphi_\Phi,\varphi_S, T)$ denote the degrees of the freedom and the field-dependent masses for particles $i$, respectively.
 We consider loop contributions from the fields $i=h^0, s^0, a_s^0, G^{\pm}, G^{0}, W_{T, L}^{\pm}, Z_{T, L}^{}, \gamma_{T, L}^{}, t$ and $b$.
 We take the $\overline{\rm MS}$ scheme, where the numerical constants $c_i$ are set at $3/2$ ($5/6$) for scalars and fermions (gauge bosons).
 The contribution of the finite temperature is defined by $I_{B,F}(a^2)= \int^{\infty}_0 dx \ x^2 \ln \left[1 \mp \exp \L-\sqrt{x^2+a^2} \R \right]$ for boson and fermions, respectively.
 The thermally corrected field-dependent masses for the CP-even/odd, Goldstone, the weak gauge bosons and top quarks are given by, for example, Ref.~\cite{Kang:2017mkl,Hashino:2016rvx}.

%%%%%%%%%%%%%%%%%%%%%%%%%%%%
%%%%%%%%%%%%%%%%%%%%%%%%%%%%
%%%%%%%%  sec: phase transition   %%%%%%%%
%%%%%%%%%%%%%%%%%%%%%%%%%%%%
%%%%%%%%%%%%%%%%%%%%%%%%%%%%
\section{Multi-step phase transitions with first order electroweak phase transition}\label{sec:pt}

 For a given scalar potential $V_{\eff}(\vec\varphi, T)$ with $\vec\varphi$ denoting a vector of real scalar fields in the multi dimensional fields space, the (critical) bubble can be found by extremizing the Euclidean action $S_E(T) \equiv S_3(T)/T$ where $S_3 (T)$ is defined as $S_3 (T) \equiv \int d^3x\left[(\partial\vec\varphi)^2/2+V_{\eff}(\vec\varphi, T)\right]$.
 Then, the bubble  nucleation rate per unit volume per unit time will be given by $\Gamma(t)=\Gamma_0(t)\exp[-S_E(t)]$ with the pre-factor $\Gamma_0 \sim T^4$.
 In order for the nucleated vacuum bubbles to percolate through the whole Universe, the nucleation rate per Hubble volume per Hubble time should reach the unity $\Gamma/H^4|_{T=T_*} \simeq 1$, which determines the transition temperature $T_*$.

 The GW spectrum from first order phase transition (FOPT) can be parameterized by several parameters, with the most crucial two, $\alpha$ and $\beta$, which capture the main features of FOPT dynamics and largely determine the features of GW spectrum.
 We will follow the conventions in Ref.~\cite{Caprini:2015zlo}.
 The parameter $\alpha\equiv \epsilon/\rho_{\rm rad}$ is the total energy budget of FOPT normalized by the radiative energy $\rho_{\rm rad}=(\pi^2/30) g_* T_*^4$ with $g_*(=108.75)$ being the relativistic degrees of freedom in the plasma at the PT temperature $T_*$.
 The liberated latent heat $\epsilon=-(\Delta V+T\partial V/\partial T)|_{T_*}$, with $\Delta V$ the vacuum energy gap between two vacua. 
 Another parameter $\beta$ is defined by $\beta \equiv - dS_E/dt|_{t_*}$.
 We use the dimensionless parameter $\wt\beta \equiv \beta/H_* $, where $H_*\equiv 1.66 \sqrt{g_*} \, T_*^2/m_{\rm pl}$ is the Hubble constant.

%%%%%%%%%%%%%%%%%%%%%%%%%%%%
%%%%%%%%%%%%%%%%%%%%%%%%%%%%
%%%%%%  sec: Numerical samples  %%%%%%
%%%%%%%%%%%%%%%%%%%%%%%%%%%%
%%%%%%%%%%%%%%%%%%%%%%%%%%%%
\section{Numerical results}\label{sec:numerical}

%%%%%%%%%%%%%%%%%%%%%%%%%%%%
%%%%%%  subsec: param  %%%%%%
%%%%%%%%%%%%%%%%%%%%%%%%%%%%
\subsection{Parameter space with various transition pattern}\label{subsec:param}
\vspace{-1mm}

 In order to study the vacuum structure at finite temperature, we use the code {\tt cosmoTransitions}~\cite{Wainwright:2011kj} for numerical studies on PT in the $Z_3$ symmetric scalar Higgs sector.
 Each path of the transition pattern and the metastable vacua at the intermediate stage of the model are shown in Fig.~\ref{fig:saddle}. 
 At $T=0$, we are interested in the case where the EWSB but $Z_3$-preserving vacuum $\Omega_h\equiv(\langle h\rangle=v, 0)$ is the ground state, which may be accompanied by a metastable vacuum $\Omega_s\equiv(0, \langle s\rangle\neq0)$ or $\Omega_{sh}\equiv(\langle h\rangle\neq0, \langle s\rangle\neq0)$.
 The presence of $\Omega_{sh}$ is a new aspect in the $Z_3$-symmetric model compared to the $Z_2$-symmetric model, and it will make possible three-step PT's in our model. 

 We summarize the parameter region of multi-step PT in Fig.~\ref{fig:As-lsh}, where two-step PT and three-step PT are plotted~\footnote{
 The one-step EWPT ($\Omega_0\ra\Omega_{h}$) is the second order for the range in Fig.~\ref{fig:As-lsh}.
 The one-step FOEWPT is realized for $m_s \gtrsim 400~\GeV$ with large $\ld_{sh}$ by the non-decoupling thermal loop effects even for $A_s=0$ as discussed in Refs.~\cite{Curtin:2014jma,Kakizaki:2015wua,Hashino:2016rvx,Beniwal:2017eik,Kurup:2017dzf}.
}.
 In the $\mu_s^2>0$ region, we find that the two-step PT ($\Omega_0\ra\Omega_s\ra\Omega_{h}$) can happen, with the first-step either second or first order, depending on the relevant parameters. 
 
  Two step (second order - first oder) PT case is basically corresponding to the $Z_2$-symmetric model in the $A_s\ra0$ limit.
 For the $\ld_s=1$ example, $A_s$ is restricted to be smaller than tens of GeV and thus the resulting deviations as expected are not significant.
 But it can still increase or decrease $T_h^*$ with appreciate amount, see the green and blue points in Fig.~\ref{fig:region} (left).
 
 Two step (first order - first oder) PT for finite $A_s$, the first-step PT significantly becomes the FOPT.
 For a large $\ld_{s}=3$, the metastable $\Omega_s$ can be accommodated for much larger $A_s\sim{\cal O}(100)~\GeV$.
 That large $A_s$, by contrast, is able to change the nature of transition $\Omega_0\ra\Omega_{s}$, into the first order type; furthermore, the strength of the second-step can be significantly enhanced and then reopens the smaller $\ld_{sh}$ region with $\ld_{sh} \sim {\cal O} (0.1)$; see Fig.~\ref{fig:region} (middle).
 We can find that the requirement $T_s^*\gtrsim T_h$ yields an upper bound on $|A_s|\lesssim300~\GeV$ in this example.  
 Note that the figures indicate that for a given $A_s$, the region for $\ld_{sh}$ is restricted and within this region increasing $\ld_{sh}$ could lead to lower $T_{h}^*$.

 The three-step (first order - second order- first oder) PT ($\Omega_0\ra\Omega_s\ra\Omega_{sh}\ra\Omega_{h}$) cases are shown in Fig.~\ref{fig:As-lsh}~(left) for $\mu_s^2<0$ and Fig.~\ref{fig:As-lsh}~(right) for $\mu_s^2>0$.
 In Fig.~\ref{fig:region} (right), we display the allowed region for $\ld_{sh} =0.24$ by taking the feasible values of $(\ld_s, m_s, A_s)$ in which we can find a point of Fig.~\ref{fig:As-lsh}~(left).
 Increasing $\ld_{s}$ lowers $T_s^*$ and it will eventually go below $T_h$, thus shutting down the three-step PT. 
 On the other hand, when $\ld_{s}$ becomes fairly small (thus for a much larger $v_s$), then $T_s^*$ ($T_h^*$) is getting higher (smaller), FOPT is enhanced in this limit.

\vspace{-2mm}
%%%%%%%%%%%%%%%%%%%%%%%%%%%%
%%%%%%%%  subsec: GW detectors %%%%%%%%
%%%%%%%%%%%%%%%%%%%%%%%%%%%%
\subsection{Detectability of gravitational waves in the $Z_3$-symmetric model}\label{subsec:prospect}
\vspace{-1mm}

\begin{figure}
 \begin{center}
\includegraphics[scale=0.4]{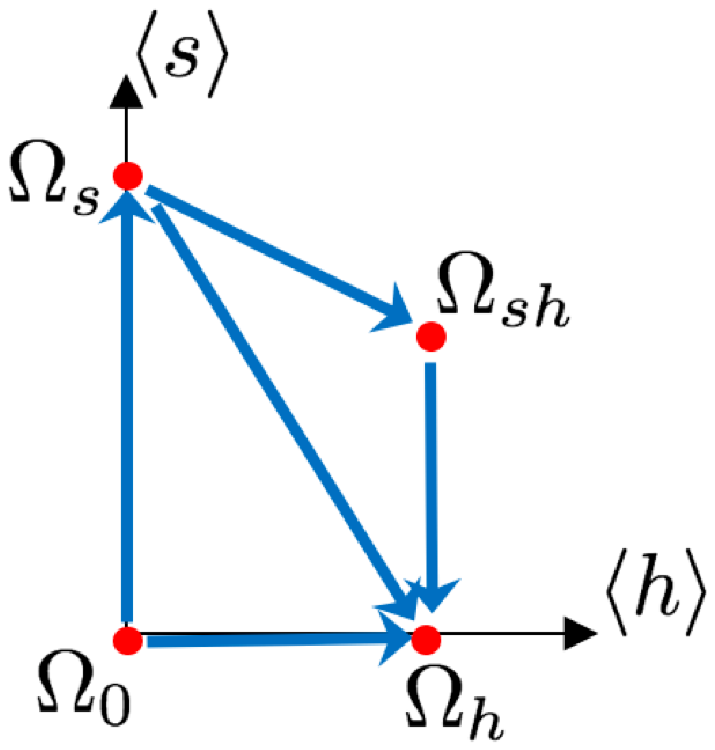}
\includegraphics[scale=0.5]{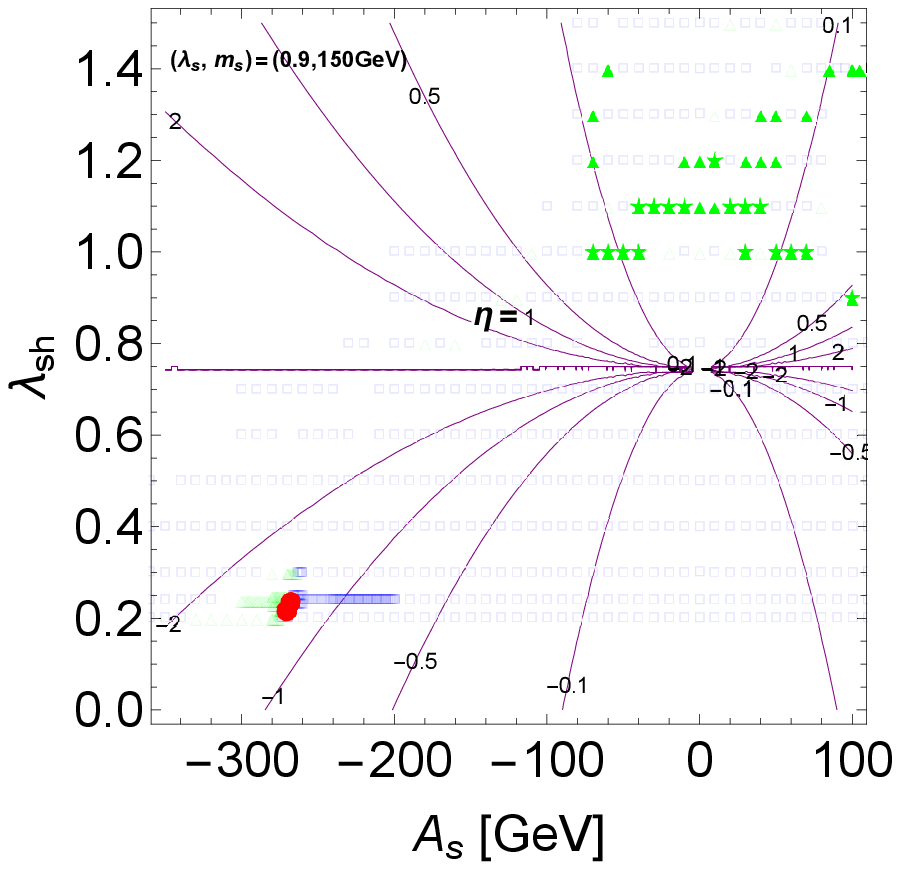}
\includegraphics[scale=0.5]{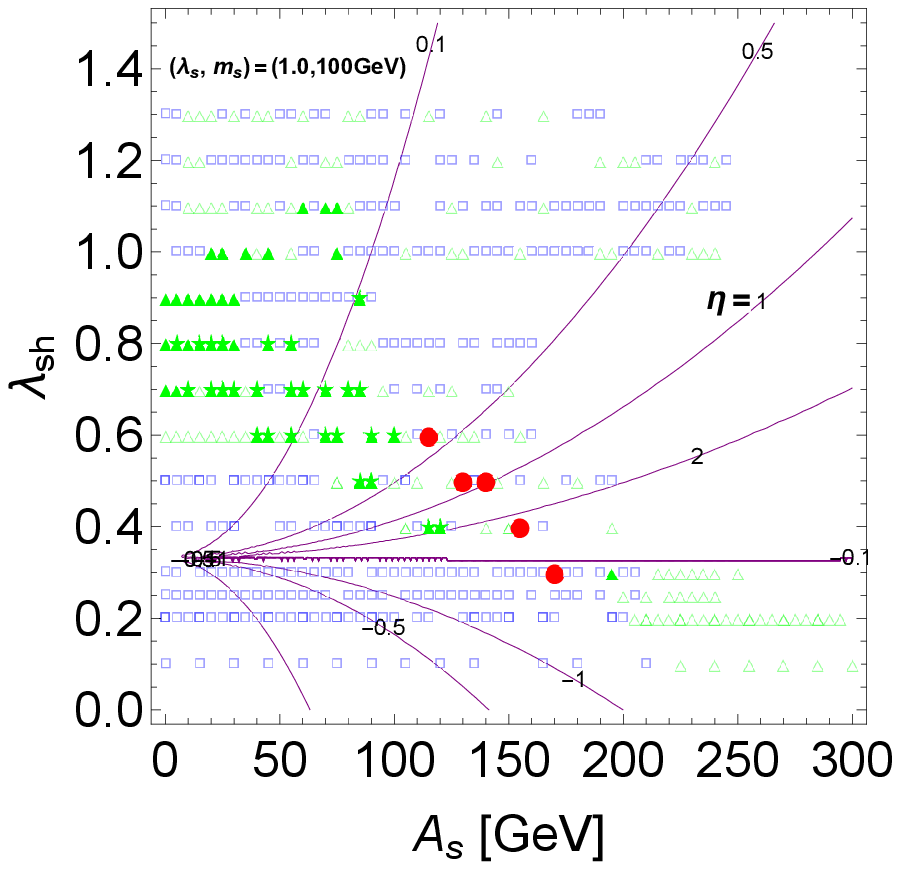}
\vspace{-3mm}
  \caption{\label{fig:saddle}\label{fig:As-lsh}
Each path of the transition pattern and the metastable vacua at the intermediate stage in the $Z_3$ model (left).
Global picture of multi-step PT in the $(A_s, \ld_{sh})$ plane for $(\ld_{s}, m_s [\GeV])=(0.9, 150)$ (left) and $(1.0, 100)$ (middle).
 PT of three-step (red, circle), two-step (green, triangle for the second-first order PT or star for the first-first order PT) and one-step (blue, square) are plotted.
 Filled plots satisfy the condition of SFOEWPT in Eq.~(\ref{eq:SFOEWPT}).
 In $\mu_s^2<0$ region, the three-step PT can happen only in a very narrow space, consistent with Fig.~\ref{fig:region} (right).
}
  \includegraphics[scale=0.44]{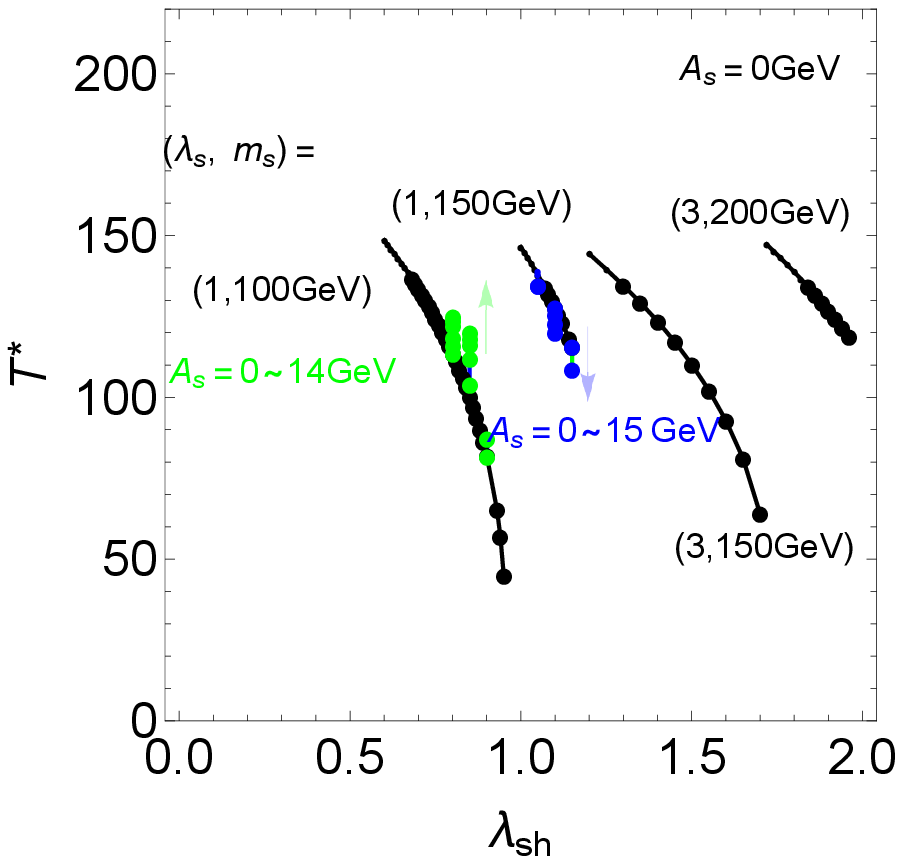}
 \includegraphics[scale=0.44]{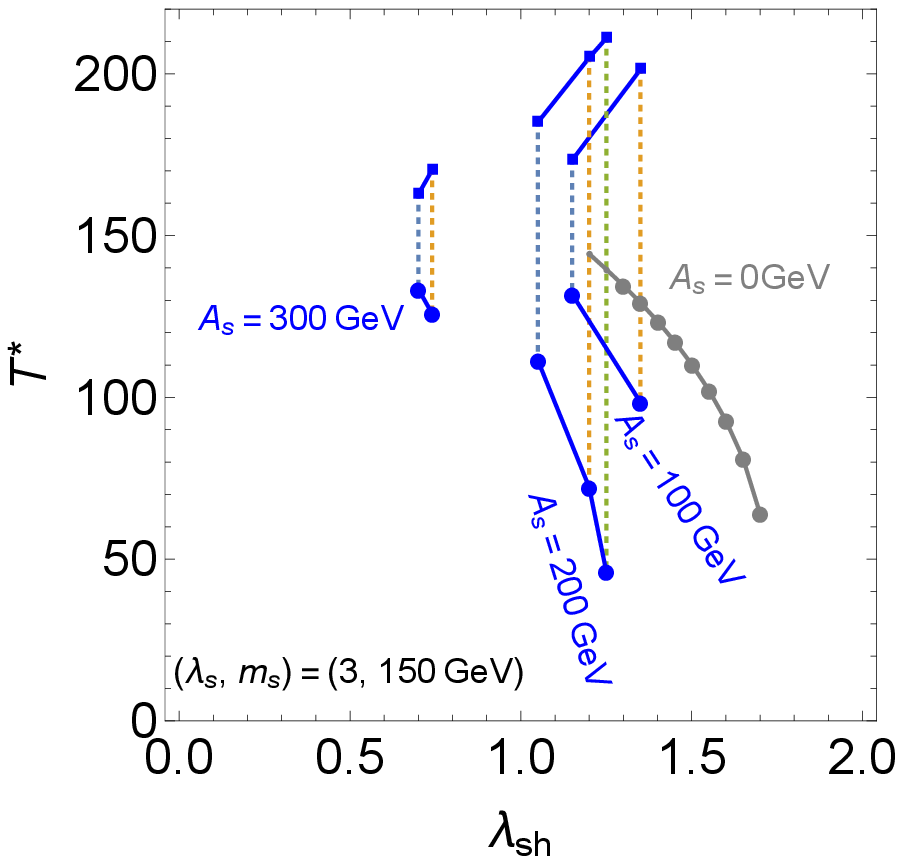}
  \includegraphics[scale=0.46]{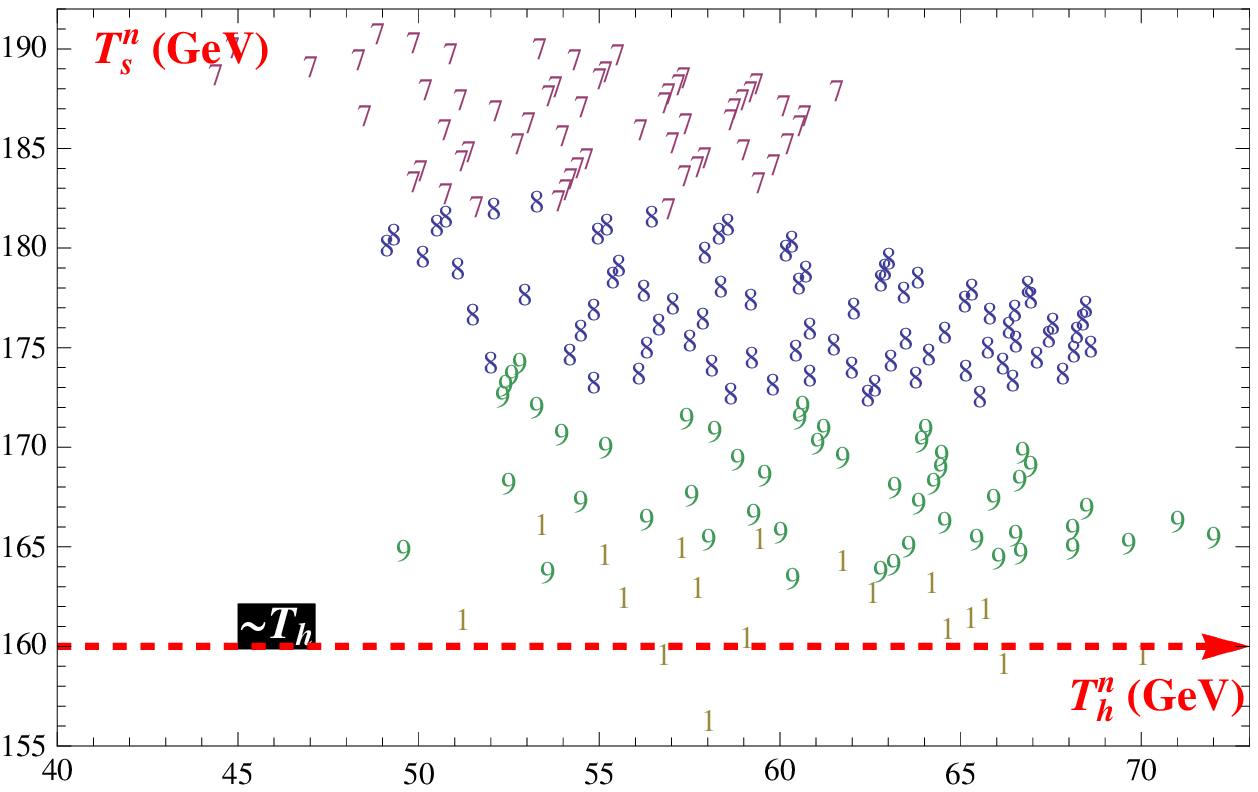}
  \vspace{-3mm}
 \caption{\label{fig:region}
%%%%%% 2-step
 (Left/middle) The two-step PT in the $\mu_s^2>0$ region as the function of $\ld_{sh}$.
 For the second-first order PT (left), we show $A_s=0$ ($Z_2$-like) case (black lines) and $A_s\neq 0$ cases by fixing $\ld_{sh}$ (green and blue lines), for four cases $(\ld_s, m_s [\GeV])=(1, 100), (1, 150), (3, 100), (3, 150)$.
 The first-first order PT (middle) arises are shon for $(\ld_s, m_s [\GeV])=(3, 150)$ by taking $A_s [\GeV]=100, 200, 300$ (blue lines).
 For each dashed line, the upper and the lower ends denote $T_s^*$ and $T_h^*$, respectively. 
 In these plots we just keep the points which give FOPT.
 %%%%% 3-step
(Right) The three-step EWPT in the $\mu_s^2<0$ region with $\ld_{sh}=0.24$, varying $\ld_s=0.7, 0.8, 0.9 ,1.0$ which is labelled by numbers 7, 8..., respectively.
 Distributions of two FOPT temperatures, $T_s^*$ and $T_h^*$; the red dashed line denotes $T_h$, the typical second order PT temperature for $\Omega_0\ra \Omega_h$.
}  
\includegraphics[scale=0.51]{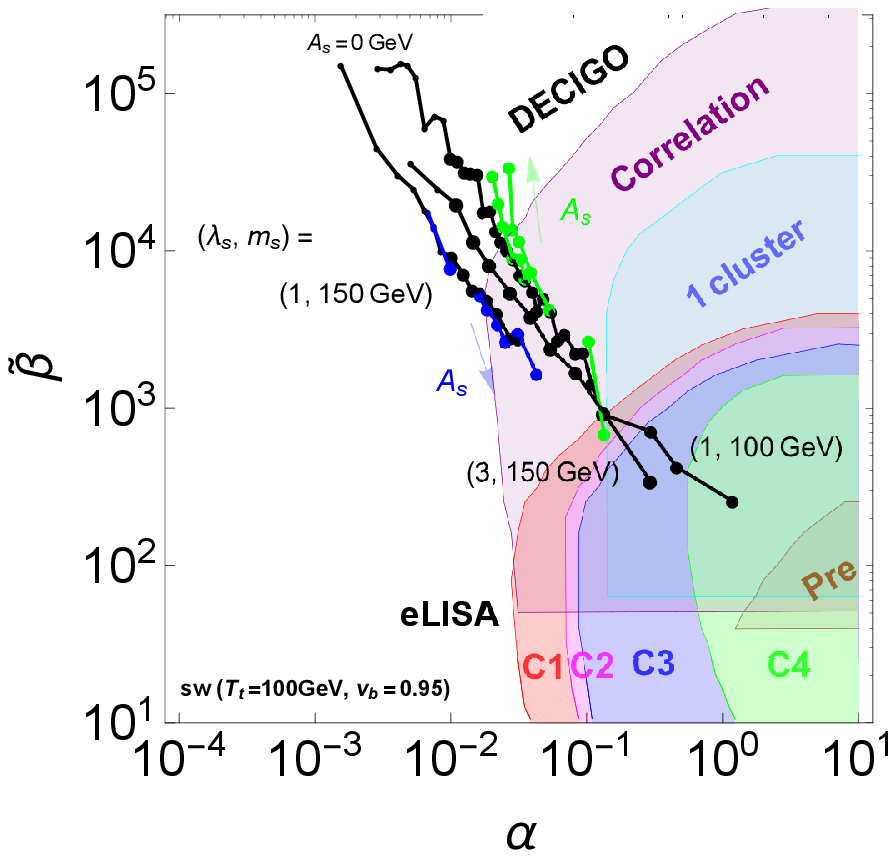}
\includegraphics[scale=0.51]{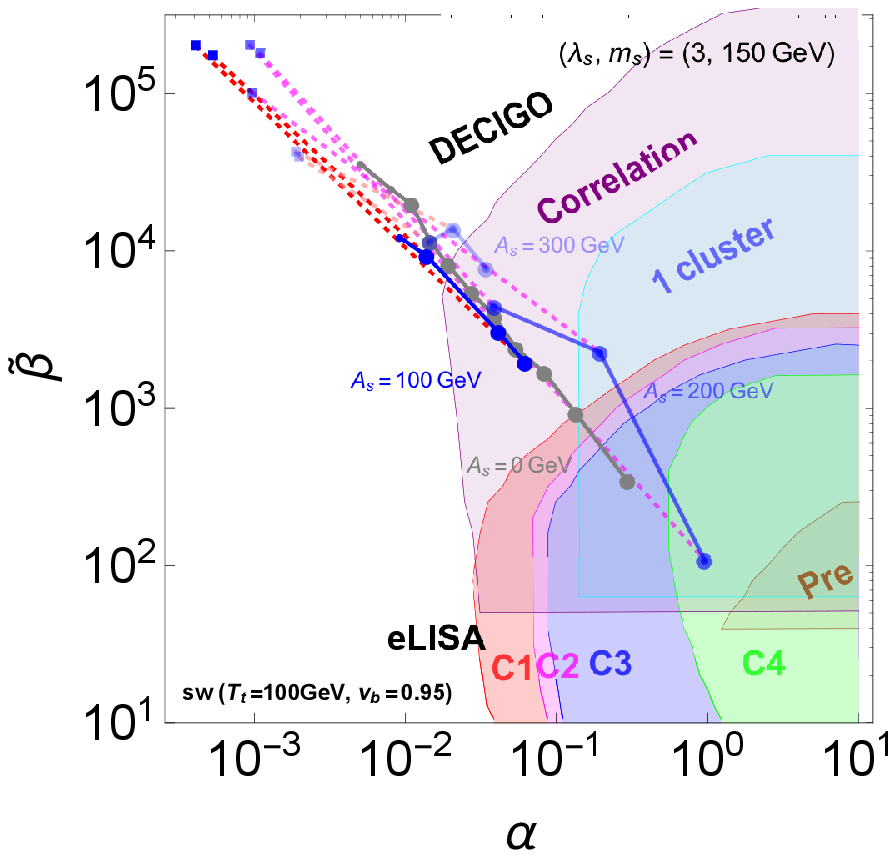}
\includegraphics[scale=0.51]{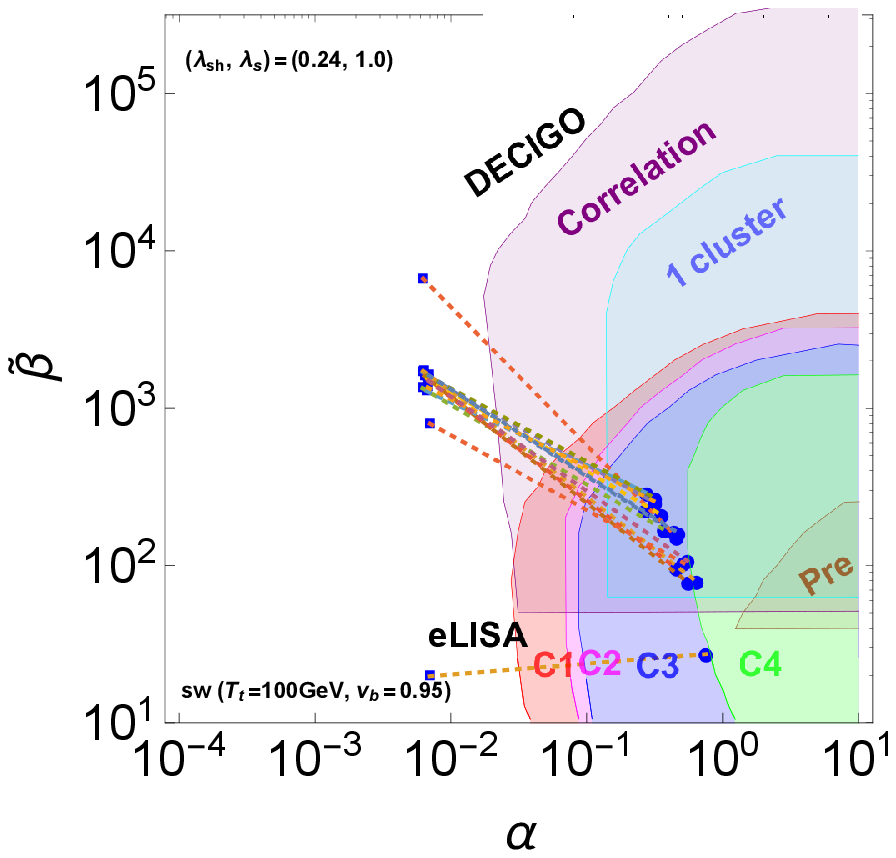}
\vspace{-3mm}
  \caption{\label{fig:a-b}
 Detectability of GWs in the ($\alpha$, $\wt\beta$) plan from the two-step (second-first/first-first order) PT (left/middle) and the three-step (first-second-first order) PT (right) which are corresponding to Fig.~\ref{fig:region}.
 In middle/left plane, the two FOPTs are labelled respectively by the square and circle points, connected by a dashed line.
 The expected sensitivities of eLISA and DECIGO are set by using the sound wave contribution for $T_*=100~\GeV$ and $v_b=0.95$. 
}
\end{center}
\end{figure}

 We display the results on the $(\alpha, \wt\beta)$ plane in the Fig.~\ref{fig:a-b}, with the experimental sensitivities of eLISA~\cite{PetiteauDataSheet,Caprini:2015zlo} and DECIGO~\cite{Kawamura:2011zz} labelled by the shaded regions. The sensitivity regions of four eLISA detector configurations described  in Table I in Ref.~\cite{Caprini:2015zlo} are denoted by ``C1'', ``C2'', ``C3'' and ``C4''.
 The expected sensitivities for the future DECIGO stages are labeled by ``Correlation'', ``1 cluster'' and ``Pre'' following Ref.~\cite{Kawamura:2011zz}.
 The transition temperature $T_*^{}$ depends on the model parameters (see, Fig.~\ref{fig:region}) and the velocity of the bubble wall $v_b^{}$ is uncertain.
 Although the experimental sensitivities on the $(\alpha, \wt\beta)$ depend on $T_*$ and $v_b^{}$, we take $T_*^{}=50~\GeV$ and $v_b=0.95$ as a reference for the purpose of illustration.
 It is seen that typically one needs $\alpha\gtrsim {\cal O}(0.01)$ for the near future detection.
 
 However, the first source from $\Omega_0\ra\Omega_s$ with FOPT turns out to be undetectable since it always gives $\alpha\lesssim 0.01$.
 On the other hand, in particular in the three-step PT case, most of the parameter space can be covered for the other source of EWPT.
 One of the main reasons causing this difference is that the first-step happened at a relatively high temperature $T_s^*\gtrsim 160~\GeV$, which typically is rather higher than the EWPT temperature $T_h^*\lesssim 100~\GeV$; recalling that $\alpha\propto 1/T^4$, thus the first source is suppressed.
 A lower $T_h^*$ also leads to smaller $\wt\beta$, which is determined by the PT temperature. 
%
%%----- figure >>>>>
%%<<<<< figure -----

%%%%%%%%%%%%%%%%%%%%%%%%%%%%
%%%%%%%%%%%%%%%%%%%%%%%%%%%%
%%%%%  sec: conclusion %%%%%
%%%%%%%%%%%%%%%%%%%%%%%%%%%%
%%%%%%%%%%%%%%%%%%%%%%%%%%%%
\section{Conclusion}\label{sec:conclusion}

 A potential barrier can be created during EWPT by the tree level effects due to a doublet-singlet mixing~\cite{Profumo:2007wc,Ashoorioon:2009nf,2step1,Fuyuto:2014yia,Profumo:2014opa,Huang:2016cjm,Hashino:2016xoj}. 
 As a result, such models can be tested by the synergy between the measurements of various Higgs boson couplings at future collider experiments and the observation of GWs at future space-based interferometers as discussed in Refs.~\cite{Huang:2016cjm,Hashino:2016xoj}.
 In another implementation imposing unbroken discrete symmetry like $Z_2$~\cite{2step1,2step2,Curtin:2014jma,Vaskonen:2016yiu,Beniwal:2017eik,Kurup:2017dzf}, multi-step PT could utilize a tree level barrier.
 But generically the absence of mixing renders the tests at colliders difficult without taking enough large $\ld_{sh}$ coupling as discussed in Refs.~\cite{Curtin:2014jma,Kakizaki:2015wua,Hashino:2016rvx,Beniwal:2017eik,Kurup:2017dzf}.
 In this paper, we have focused on such the nightmare scenario in the $Z_3$ symmetric single scalar model.
 Especially, the three-step PT produces two sources of GW in the model.
 Despite of the undetectability from the first-step in the near future, the other source from EWPT basically can be completely covered by LISA and DECIGO.

\vspace{-2mm}
%%%%%%%%%%%%%%%%%%%%%%%%%%%%
%%%%%%%%%%%%%%%%%%%%%%%%%%%%
%%%%%%%  sec: acknowledgements %%%%%%%
%%%%%%%%%%%%%%%%%%%%%%%%%%%%
%%%%%%%%%%%%%%%%%%%%%%%%%%%%
\section*{Acknowledgements}\label{sec:acknowledgements}

 This work is based on the collaboration with Zhaofeng Kang and Pyungwon Ko. 
 I would like to thank them for their support.

\vspace{-2mm}
%%%%%%%%%%%%%%%%%%%%%%%%%%%%
%%%%%%%%%%%%%%%%%%%%%%%%%%%%
%%%%%%%%%%%  refferences %%%%%%%%%%%
%%%%%%%%%%%%%%%%%%%%%%%%%%%%
%%%%%%%%%%%%%%%%%%%%%%%%%%%%

\end{document}